\documentclass[aps,prl,twocolumn,groupedaddress,showpacs,amsmath,amssymb]{revtex4}

\usepackage[dvips]{graphicx}
\usepackage{latexsym}

\begin{document}


\title{Nonequilibrium Steady State of Photoexcited Correlated Electrons \\
in the Presence of Dissipation
} 

\author{Naoto Tsuji, Takashi Oka, and Hideo Aoki}
\affiliation{Department of Physics, University of Tokyo, Hongo, 
Tokyo 113-0033, Japan}


\date{\today}

\begin{abstract}
We present a framework to determine nonequilibrium 
steady states in strongly correlated electron systems 
in the presence of dissipation.  
This is demonstrated for 
a correlated electron (Falicov-Kimball) model 
attached to a heat bath and 
irradiated by an intense pump light, for which an exact solution 
is obtained 
with the Floquet method combined with the 
nonequilibrium dynamical mean-field theory. 
On top of a Drude-like peak indicative of photometallization 
as observed in recent pump-probe experiments, 
new nonequilibrium phenomena are predicted to emerge, where 
the optical conductivity exhibits 
dip and kink structures around the frequency of the 
pump light, a midgap absorption arising from photoinduced Floquet subbands, 
and a negative attenuation (gain) due to a population inversion.
\end{abstract}



\maketitle


{\it Introduction.}---Recent pump-probe experiments have unveiled
not only ultrafast dynamics of correlated electron 
systems, but also a possibility to control their physical properties and 
even trigger ``phase transitions'' with irradiation of pump lights.
A typical example is photo-induced insulator-metal transition 
\cite{Iwai2003a, Cavalleri2004a, Okamoto2007a}, 
where the time-resolved optical conductivity 
[$\sigma(\nu)$, $\nu$: the photon energy of the probe light]
exhibits a Drude-like peak in a low-energy region that indicates 
photocarriers drive the system into a metal. 
This ``photodoping'' 
opens up a new frontier for controlling the phase of correlated electron 
materials as an alternative to chemical doping.

Experimentally it is known for 
strongly correlated electron systems that an excited state relaxes 
very fast ($\lesssim$ 1 ps) 
after the pumping is turned off \cite{Okamoto2007a},
which implies that the system is subject to strong dissipation.  
Thus we expect that during irradiation of the pump light the balance between
pumping and relaxation is rapidly achieved, and a 
nonequilibrium steady state (NESS) emerges. 
To identify NESS is indeed a long-standing issue in nonequilibrium quantum statistical mechanics. 
For many-body electron problems, NESS has been mainly studied for quantum dot systems
(e.g., Ref.~\onlinecite{MeirWingreenLee1993}),
while we are still some way from a firm understanding of NESS for bulk systems 
that might exhibit phase transitions.  
On the other hand, the relevant dissipation mechanism is still not clear.
Time evolution of isolated correlated electron systems
has been 
studied for photoexcited one-dimensional systems
\cite{photoinduced1D}
or for the nonequilibrium Falicov-Kimball (FK) model 
\cite{FreericksTurkowskiZlatic2006, EcksteinKollar2008b}.  
These studies, however, do not elaborate the role of {\it extrinsic}
dissipation that can be relevant to
the long-time steady-state behavior \cite{dissipation}.

Motivated by these, we pose two issues here: 
(i) can we present a general argument for 
NESS determined by dissipation in a photoirradiated electron system, 
and (ii) are there novel phenomena that emerge specifically 
in nonequilibrium?  
For the former we have performed an 
exact analysis of NESS for the FK model in the framework of
the Floquet technique \cite{Shirley1965, Sambe1973} as combined with 
the dynamical mean-field theory (DMFT) \cite{GeorgesKotliarKrauthRozenberg1996a}.
For the latter, our new finding in the observable $\sigma(\nu)$ characteristic to NESS is 
dip and kink structures around $\nu \sim \Omega$ 
(where $\Omega$ is the frequency of the pump light), 
a midgap peak arising from photoinduced Floquet subbands, 
and a negative peak (gain) arising from a population inversion.
These predictions offer experimental opportunities
to realize and control the novel state with photoirradiation.

{\it Model}.---In order to include dissipation, let us consider a system coupled to a heat bath
with the total Hamiltonian
\begin{align}
  H_{\rm tot}
	  &=
		  H_{\rm syst} + H_{\rm mix} + H_{\rm bath}.
	\label{hamiltonian}
\end{align}
As a solvable model for the heat bath, we take the ``B\"{u}ttiker probe'' reservoir 
\cite{Buttiker1985a}, 
\begin{align}
	H_{\rm mix}
	  &=
		  \sum_i \sum_p
			\left[V_p (c_i^\dagger b_{i,p} + b_{i,p}^\dagger c_i) 
			+\frac{V_p^2}{\varepsilon_p}c_i^\dagger c_i\right],
	\label{bath1}
	\\
	H_{\rm bath}
	  &=
		  \sum_i \sum_p \varepsilon_p b_{i,p}^\dagger b_{i,p},
	\label{bath2}
\end{align}
where $c_i^\dagger$ ($b_{i,p}^\dagger$) creates an electron 
(bath's fermionic degrees of freedom), $\varepsilon_p$ is the 
kinetic energy, and $V_p$ is
the coupling to the mode $p$ of the bath. 
The thermal bath is in equilibrium with temperature $T$.
Its chemical potential is determined so that the current does not flow 
between the bath and the system.
In $H_{\rm mix}$  we have included a contour term 
[the second term on the right hand side of Eq. (\ref{bath1})]
that cancels the potential shift due to 
the coupling to the bath.  
Let us now define a hybridization function 
$\Gamma(\omega)=\sum_p \pi V_p^2 \delta(\omega-\varepsilon_p)$, 
and the system dissipates with a damping rate $\Gamma$.  
For simplicity, we omit
the $\omega$ dependence of $\Gamma$.

When one starts driving the system with a continuous pump light
at time $t=0$, a transient dynamics 
from the initial equilibrium state to an excited state should first 
occur, but then 
the system will relax to an NESS within $t \sim \Gamma^{-1}$.
Here we assume that (i) NESS exists, and (ii)
NESS should be independent of the initial 
condition and correlations between the initial state and NESS, 
since we expect that those effects
will be wiped out in the presence of dissipation (while 
in the absence of dissipation this is shown to not necessarily be
the case \cite{Tran2008}).

Based on these assumptions, NESS is determined as follows.
The absence of the initial correlations allows us to use the Keldysh formalism.
Since the pump light is an ac electric field periodic in time, 
NESS is also time-periodic, so that we can employ 
the Floquet method. 
After integrating out the bath degrees of freedom, 
we obtain the Dyson equation in a Floquet matrix form \cite{TsujiOkaAoki2008}
denoted by hats,
\begin{align}
  &
  \begin{pmatrix}
	  \hat{G}_{\boldsymbol k}^R(\omega) & \hat{G}_{\boldsymbol k}^K(\omega) \\
		0 & \hat{G}_{\boldsymbol k}^A(\omega)
	\end{pmatrix}^{-1}
	  =
		  \begin{pmatrix}
			  (\hat{G}_{\boldsymbol k}^{R0})^{-1}(\omega) & (\hat{G}_{\boldsymbol k}^{-1})^{K0}(\omega) \\
				0 & (\hat{G}_{\boldsymbol k}^{A0})^{-1}(\omega)
			\end{pmatrix}
	\nonumber
	\\
    &\qquad+
		  \begin{pmatrix}
			  i\Gamma\, \hat{1} & 2i\Gamma\hat{F}(\omega)\\
				0 & -i\Gamma\, \hat{1}
			\end{pmatrix}
	  -
			\begin{pmatrix}
			  \hat{\Sigma}^R(\omega) & \hat{\Sigma}^K(\omega) \\
				0 & \hat{\Sigma}^A(\omega)
			\end{pmatrix},
	\label{dyson}
\end{align}
where $\hat{G}^{R,A,K}$($\hat{G}^{R0,A0,K0}$) are 
respectively full (noninteracting)
retarded, advanced, and Keldysh Green's functions, 
$\hat{\Sigma}^{R,A,K}$ are respective self-energies, 
$\hat{1}$ the identity matrix, and 
$(\hat{F})_{mn}(\omega)=\delta_{mn} \tanh[(\omega+n\Omega)/2T]$.
In Eq.~(\ref{dyson}), 
$(\hat{G}^{R0\;-1}_{\boldsymbol k})_{mn}(\omega)=
(\omega+n\Omega+\mu+i\eta)\delta_{mn}-(\hat{\epsilon}_{\boldsymbol k})_{mn}$
with $\mu$ the chemical potential, $\eta$ a positive infinitesimal, 
and $\hat{\epsilon}_{\boldsymbol k}$ the Floquet matrix for 
the noninteracting Hamiltonian \cite{TsujiOkaAoki2008}. 
The Keldysh component $(\hat{G}^{-1})^{K0}$ in Eq.~(\ref{dyson}),
on the other hand, vanishes since it is proportional to $i\eta$ 
while there is a nonzero dissipation term $2i\Gamma\hat{F}$.
This means that, although $(\hat{G}^{-1})^{K0}$ contains information 
on the initial condition before the 
ac field is applied as well as on the way in which the field 
is turned on, it is again wiped out due to dissipation. 
NESS is thus determined without ambiguity by the Dyson equation ~(\ref{dyson}).
We note that if we stand on a phenomenological point of view, we are not restricted to the 
specific model of the reservoir (\ref{bath1}) and (\ref{bath2}). Alternatively, 
we can start from Eq.~(\ref{dyson}), interpreting $\Gamma^{-1}$
as the relaxation time for a relevant dissipation mechanism such as phonons or spins.
%

\begin{figure}[t]
  \begin{center}
    \includegraphics[width=7cm]{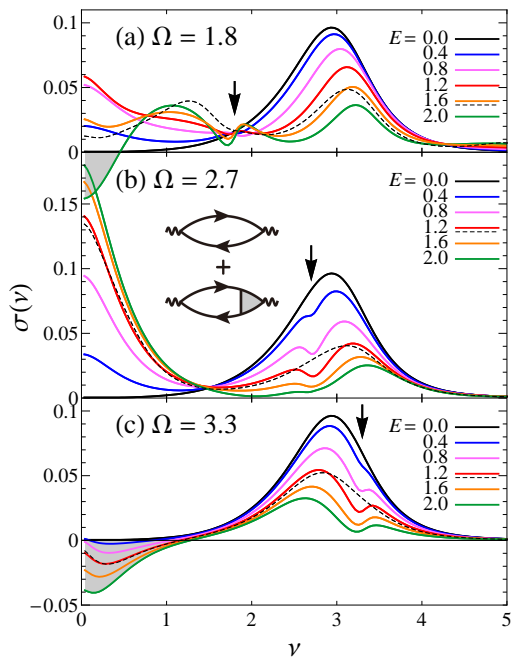}
		\caption{(color online) Real part of the optical conductivity $\sigma(\nu)$ (solid lines) 
		in units of $\sigma_0$ for (a) $\Omega=1.8$, (b) $2.7$ 
		and (c) $3.3$. 
		The dashed lines illustrate (for specific values of $E$) the results without the vertex correction. 
		The arrows indicate the frequency $\Omega$ of the pump light. 
		Inset in (b) depicts diagrams of the bubble and the vertex correction.}
		\label{sigma}
	\end{center}
\end{figure}

As one of the simplest models of correlated electron 
systems we take the FK model \cite{FalicovKimball1969, FreericksZlatic2003}, 
\begin{align}
  H_{\rm syst}
	  &=
		  \sum_{\boldsymbol k} \epsilon_{{\boldsymbol k}-e{\boldsymbol A}(t)}
			c_{\boldsymbol k}^\dagger c_{\boldsymbol k}
			+ U \sum_i c_i^\dagger c_i f_i^\dagger f_i,
	\label{fk}
\end{align}
irradiated with a uniform pump light with the vector potential 
${\boldsymbol A}(t)$, where 
$\epsilon_{\boldsymbol k}$ is the energy dispersion of a single-band system, 
$U$ is the interaction strength, and 
$f_i^\dagger$ creates a localized electron.  
This model is known \cite{FreericksZlatic2003} to undergo a metal-insulator transition as we change $U$.
We adopt the model since 
(i) it has a simple optical excitation spectrum 
with a single charge transfer (CT) peak around $\nu = \Omega_{\rm CT} \sim U$
(Fig.~\ref{sigma}, the black curves), 
(ii) charge and spin degrees of freedom decouple from the outset 
so that we can concentrate on charge dynamics
[we drop spin indices in Eq.~(\ref{fk})], and
(iii) it can be solved {\it exactly} within DMFT
\cite{GeorgesKotliarKrauthRozenberg1996a} even out of equilibrium
\cite{FreericksTurkowskiZlatic2006}.

%

{\it Method}.---To treat periodically driven correlated electrons, 
we can make use of the Floquet method combined with DMFT
\cite{TsujiOkaAoki2008} (an equivalent numerical technique was used in 
Ref.~\onlinecite{JouraFreericksPruschke2008}). 
Here we adopt this method for the dissipative case including the 
Keldysh component using Eq.~(\ref{dyson}).
We consider the hypercubic lattice with 
$\epsilon_{\boldsymbol k}=-t^\ast \sum_{i=1}^d \cos k_i/\sqrt{d}$
in $d$ ($\to \infty$ in DMFT) dimensions. For simplicity, we assume that 
${\boldsymbol A}(t)=A(t)(1,1,\dots,1)$ with $A(t)=-E\sin\Omega t/\Omega$
(where $E$ is the amplitude of the pump light). 
Then 
$(\hat{\epsilon}_{\boldsymbol k})_{mn} = \epsilon_{\boldsymbol k} J_{m-n}(eE/\Omega)$
for $m-n$ even, or 
$i\,v_{\boldsymbol k} J_{m-n}(eE/\Omega)$
for $m-n$ odd \cite{TsujiOkaAoki2008}
with $v_{\boldsymbol k}=t^\ast \sum_{i=1}^d \sin k_i/\sqrt{d}$ and 
$J_n(z)$ the $n$th order Bessel function. 
We take $t^\ast$ as the unit of energy.
The integration over ${\boldsymbol k}$ may be done with 
the joint density of states
$\rho(\epsilon, v)=\sum_{\boldsymbol k}\delta(\epsilon-\epsilon_{\boldsymbol k})
\delta(v-v_{\boldsymbol k})=e^{-\epsilon^2-v^2}/\pi$ \cite{TurkowskiFreericks2005}.


The optical conductivity is defined
in such a way that the current change due to an infinitesimal probe light 
$\delta \mathcal{E}e^{-i\nu t}$ (which we assume to be parallel to the pump light) be
$\delta j(t)=\sigma(\nu) \delta \mathcal{E}e^{-i\nu t}+\sum_{n\neq 0}(\cdots) 
\delta\mathcal{E} e^{-i(\nu+n\Omega)t}$,
and is given by Kubo-like formula \cite{TsujiOkaAoki2009}
\begin{align}
  \sigma(\nu)
	  =&
		  \frac{\sigma_0}{\nu} \sum_{\boldsymbol k} \frac{1}{\tau} \int_{-\tau/2}^{\tau/2} d\bar{t}\;
			\bigg\{
			-\epsilon_{{\boldsymbol k}-e{\boldsymbol A}(\bar{t})} G_{\boldsymbol k}^<(\bar{t}, \bar{t})
	\nonumber
	\\
	  &+
			\int_{0}^{\infty}	dt\; e^{i\nu t}\; 
		  \Big\langle
			  [
				  j_{\boldsymbol k} (\bar{t}+t/2),\, 
					j_{\boldsymbol k} (\bar{t}-t/2)
				]
			\Big\rangle\bigg\},
	\label{optical}
\end{align}
with $\sigma_0\equiv e^2/d$, $\tau\equiv 2\pi/\Omega$, 
$G^<$ the lesser Green's function,
$[\, , \,]$ the commutator, 
$\langle \dots \rangle$ the statistical average,
and $j_{\boldsymbol k}(t)=v_{{\boldsymbol k}-e{\boldsymbol A}(t)}\,
c_{\boldsymbol k}^\dagger(t) c_{\boldsymbol k}(t)$.
In the limit 
$E\to 0$, 
we reproduce the equilibrium linear-response theory.

In calculating $\sigma(\nu)$, we divide Eq.~(\ref{optical}) into two parts: 
the bubble diagram and the vertex correction [see the inset of Fig.~\ref{sigma} (b)]. 
The former for ${\rm Re}\;\sigma(\nu)$ is written as 
\begin{align}
		& \frac{\pi\sigma_0}{\nu}
			\sum_{\boldsymbol k} \int_{-\Omega/2}^{\Omega/2} d\omega\;
			{\rm Tr}[
				  \hat{v}_{\boldsymbol k} \hat{A}_{\boldsymbol k}(\omega+\nu) 
				  \hat{v}_{\boldsymbol k} \hat{N}_{\boldsymbol k}(\omega)
	\nonumber
	\\
	  &\quad
				- \hat{v}_{\boldsymbol k} \hat{A}_{\boldsymbol k}(\omega) 
				  \hat{v}_{\boldsymbol k} \hat{N}_{\boldsymbol k}(\omega+\nu)
			],
	\label{bubble}
\end{align}
where $\hat{A}_{\boldsymbol k}(\omega)=
i[\hat{G}_{\boldsymbol k}^R(\omega)-\hat{G}_{\boldsymbol k}^A(\omega)]/2\pi$, 
$\hat{N}_{\boldsymbol k}(\omega)=-i\hat{G}_{\boldsymbol k}^<(\omega)/2\pi$, 
the trace runs over the Floquet indices, and the bare vertex function
$(\hat{v}_{\boldsymbol k})_{mn} = v_{\boldsymbol k} J_{m-n}(eE/\Omega)$ 
for $m-n$ even, or 
$-i\,\epsilon_{\boldsymbol k} J_{m-n}(eE/\Omega)$
for $m-n$ odd.

The vertex correction to $\sigma(\nu)$, on the other hand, 
exactly vanishes in equilibrium within DMFT due to the odd parity of $v_{\boldsymbol k}$. 
However, this no longer holds for nonequilibrium cases, since
the bare vertex $v_{{\boldsymbol k}-e{\boldsymbol A}(t)}$ is generally not parity odd. 
We note that 
the vertex correction strongly depends on the relative polarization
of the pump and probe lights.
A virtue of the Floquet formalism is that the vertex correction 
can be calculated exactly 
with the self-consistent equations for dressed vertex functions \cite{TsujiOkaAoki2009}. 
In practical calculations, we introduce a cutoff for the Floquet matrix size 
(typically 7), for which we have checked the error is small enough. 
In the following we put $\Gamma=0.05, T=0.05$, and $U=3$
(so the system is insulating),
and set the system at half-filling for both $c$ and $f$ electrons.

{\it Numerical results}.---We show the results for $\sigma(\nu)$ 
in Fig.~\ref{sigma} for three cases: 
(a) the frequency of the pump light $\Omega=1.8 < \Omega_{\rm CT} \sim U=3$, 
(b) $2.7\lesssim\Omega_{\rm CT}$, and (c) $3.3 \gtrsim \Omega_{\rm CT}$. 
When the pump light is absent ($E=0$, the black curves in Fig.~\ref{sigma}), an optical gap 
is clearly seen in the low-energy region. 
As we increase the amplitude $E$, we find that CT peak ($\nu \sim \Omega_{\rm CT}$)
collapses in all the three cases, which is due to the bleaching effect.
On the other hand, the low-energy behavior of $\sigma(\nu)$ is dramatically 
different for the three cases: 
In Fig.~\ref{sigma}(b), a positive peak appears around $\nu \sim 0$, 
which implies that the system is driven into a metallic state.
Strikingly, in Fig.~\ref{sigma}(c) we find 
a {\it negative weight} (hatched in Fig.~\ref{sigma}), 
which suggests that the system has some energy gain.
In the case of Fig.~\ref{sigma}(a) where one-photon process
is forbidden due to $\Omega < \Omega_{\rm CT} \sim U$, 
we can notice a midgap absorption around $\nu \sim 1.2$ 
with a positive or negative weight in $\nu \sim 0$.

\begin{figure}[t]
  \begin{center}
    \includegraphics[width=8cm]{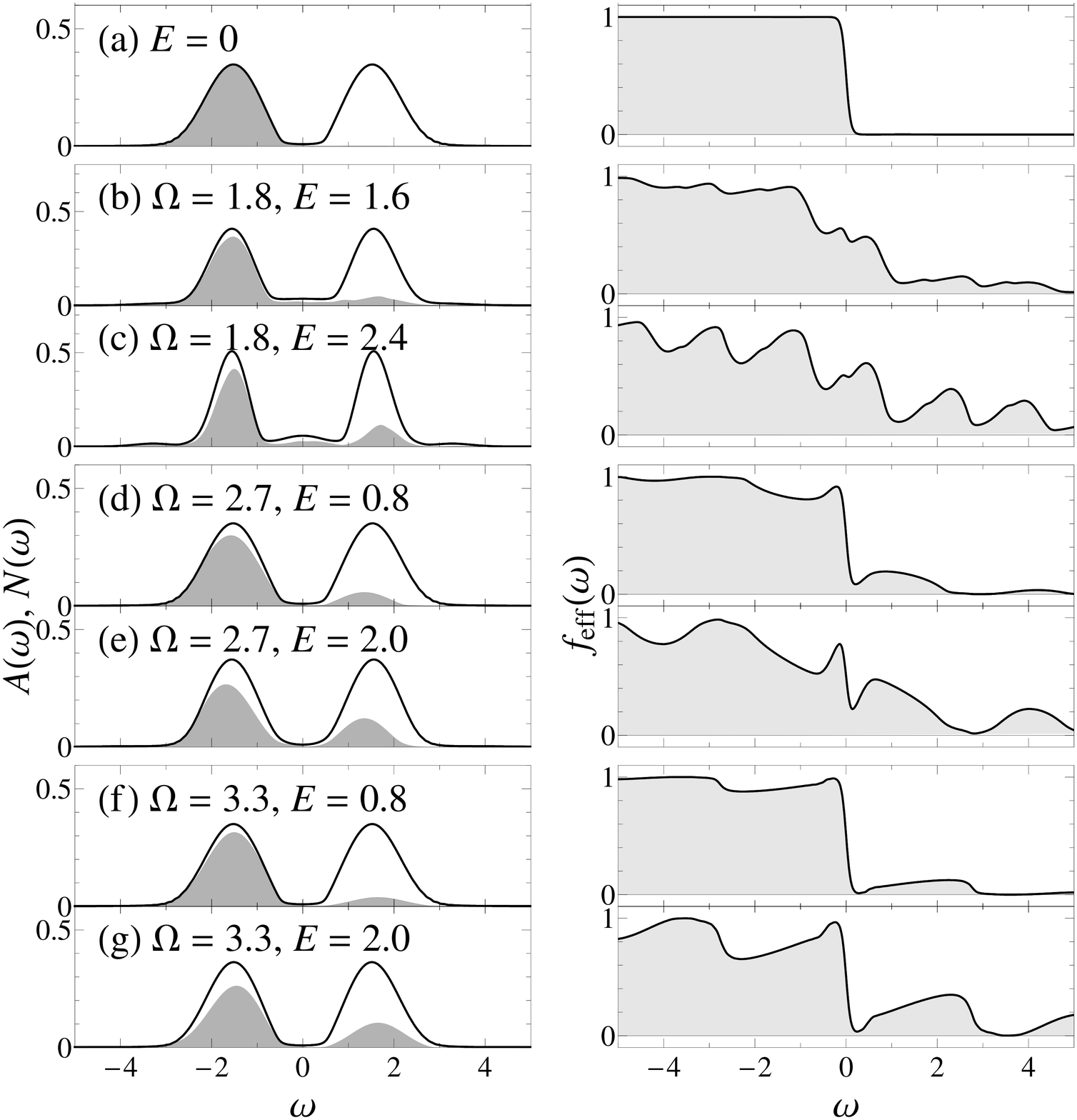}
		\caption{Left: Time-averaged density of states $A(\omega)$ (solid curves) and 
		time-averaged density of occupied states $N(\omega)$ (shaded regions) 
		for (a) $E=0$, (b),(c) $\Omega=1.8$, (d),(e) $\Omega=2.7$, and (f),(g) 
		$\Omega=3.3$ with 
		several $E$. 
		Right: The corresponding effective distributions $f_{\rm eff}(\omega)=N(\omega)/A(\omega)$.}
		\label{spec_f}
	\end{center}
\end{figure}
We also find that there are new features, besides the 
low-energy behavior, appearing 
around the pump frequency $\nu \sim \Omega$ in all the three cases: 
a {\it kink} in Fig.~\ref{sigma}(a) and {\it dips} in Figs.~\ref{sigma}(b) and \ref{sigma}(c).
These features are unique to NESS, and are not seen in equilibrium.  
To reveal the origins of the newly found features, 
let us first clarify the effect of the vertex correction by comparing the 
results with and without the correction 
(the solid lines and dashed lines in Fig.~\ref{sigma}, respectively).
We find that the correction can contribute to $\sigma(\nu)$ quite significantly 
around $\nu \sim \Omega$, creating the kink [Fig.~\ref{sigma}(a)]
and dip [Figs.~\ref{sigma}(b) and \ref{sigma}(c)] structures. 
Since there is no such correction 
in equilibrium, these structures are a genuine quantum many-body effect 
{\it unique in nonequilibrium}. 
We emphasize that the effect is 
distinct from the so-called ``hole burning'' effects 
observed in inhomogeneous systems \cite{ButcherCotter}.

Other features can be well captured by the bubble diagram (\ref{bubble}).
To analyze them, we calculate time-averaged 
local density of states (DOS) 
$A(\omega+n\Omega)=\sum_{\boldsymbol k}(\hat{A}_{\boldsymbol k})_{nn}(\omega)$, 
time-averaged density of occupied states 
$N(\omega+n\Omega)=\sum_{\boldsymbol k}(\hat{N}_{\boldsymbol k})_{nn}(\omega)$
(Fig.~\ref{spec_f}, left-hand panels), 
and an effective distribution
$f_{\rm eff}(\omega) = N(\omega)/A(\omega)$ (right-hand panels).

In the absence of the pump light [$E=0$, 
Fig.~\ref{spec_f}(a)] we can see an insulating state 
equilibrated with the Fermi distribution. 
As the pump field is turned on, 
DOS changes in different manners according to which regime $\Omega$ belongs to:
For $\Omega \sim \Omega_{\rm CT}$ DOS hardly changes [Figs.~\ref{spec_f}(d)-\ref{spec_f}(g)], 
and the energy gap remains [Figs.~\ref{spec_f}(e) and \ref{spec_f}(g)] even when the CT 
peak in $\sigma(\nu)$ almost disappears [Figs.~\ref{sigma}(b) and \ref{sigma}(c)]. 
For $\Omega < \Omega_{\rm CT}$ in Figs.~\ref{spec_f}(b) and \ref{spec_f}(c), by contrast, 
the DOS is modulated to have photoinduced midgap states, 
which we can assign to the Floquet subbands \cite{TsujiOkaAoki2008} 
(a set of replicas of the original band with a spacing $\Omega$). 
These states are responsible for the midgap peak in $\sigma(\nu)$ 
[Fig.~\ref{sigma}(a)] 
corresponding to one-photon process with absorption or emission
energy $\Omega_{\rm CT}-\Omega\sim 1.2$. 

Let us turn to the distribution (Fig.~\ref{spec_f}, right-hand panels).
It is greatly modified by the pump light, where
electrons occupying the lower band are 
pushed to the upper band absorbing the photon energy $n\Omega$, 
and become photocarriers. 
What is newly found here is that the nonequilibrium distribution 
$f_{\rm eff}(\omega)$ becomes highly {\it nonmonotonic} with 
characteristic periodic structures (the period $\sim \Omega$) as 
clearly displayed in the right-hand panels of Fig.~\ref{spec_f}. 
Since we cannot fit $f_{\rm eff}(\omega)$
to the Fermi distribution with an elevated temperature, 
this is again a genuinely nonequilibrium effect, 
where heating picture is not applicable.

\begin{figure}[t]
  \begin{center}
    \includegraphics[width=7cm]{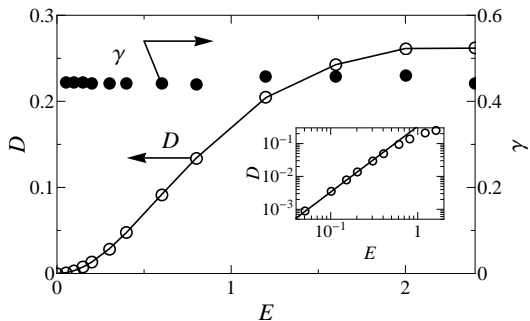}
		\caption{Crossover from an insulator to a photoinduced metal: 
		the weight $D$ (open circles, joined as a guide to the eye) 
		and width $\gamma$ (filled circles) against $E$ for $\Omega=2.7$. 
		Inset is a blowup with a log-log plot, where the 
		line is a fit, $D=0.335E^2$.}
		\label{d_gamma}
	\end{center}
\end{figure}

The low-energy behavior of $\sigma(\nu)$ can also be explained with the distribution.
For $\Omega=2.7\lesssim\Omega_{\rm CT}$, 
electrons in the upper part of the lower band tend to be excited into the lower part of the upper band 
[Figs.~\ref{spec_f}(d) and \ref{spec_f}(e)], whereas 
electrons in the lower part of the lower band tend to be excited into the upper part of the upper band 
for $\Omega=3.3\gtrsim\Omega_{\rm CT}$ 
[Figs.~\ref{spec_f}(f) and \ref{spec_f}(g)]. 
In the former distribution, 
electrons prefer to absorb infinitesimal energy rather than emit, 
which is why the photo-carriers induce a Drude-like peak 
despite the energy gap remaining in the DOS. 
In the latter, 
a {\it population inversion} within each band is realized in 
the steady state, giving 
a negative weight in $\sigma(\nu)$. This implies an energy gain, 
i.e., a transmitted light acquires intensity stronger than the incident light.   
A negative $\sigma(\nu)$ is also observed for $\Omega=1.8<\Omega_{\rm CT}$
and sufficiently large $E$ [Fig.~\ref{sigma}(c)],
which is now explained with two-photon processes, 
with absorption energy $\sim 2\Omega\gtrsim\Omega_{\rm CT}$. 

How does the nature of
the low-energy metallic peak in Fig.~\ref{sigma}(b)
evolve with the pump light amplitude $E$?  
We fit the deviation of $\sigma(\nu)$ from the 
equilibrium one,
$\delta\sigma(\nu)=\sigma(\nu)-\sigma(\nu)|_{E=0}$,
to a Drude-like expression
$\sigma_0 t^\ast D\gamma/[\pi(\nu^2+\gamma^2)]$
by least squares. 
Here the weight $D$ and the width $\gamma$ are fitting parameters, 
and the fitting is performed 
for a low frequency region ($\nu<0.8 t^\ast$, here), with 
the rms error turning out to be $<4$\%.  
The result in Fig.~\ref{d_gamma} shows that
$D$ starts to grow nonlinearly, and is saturated around $E \sim 2$.  
We do not have a phase transition with singularity, but we do have 
a crossover from an insulator to a metal. 
$\gamma$, on the other hand, is nearly constant against $E$.
This behavior is similar to the chemically doped FK model, where 
a single-particle excitation has a finite lifetime at $T=0$ 
\cite{FreericksZlatic2003},
so that 
the photoinduced state is also a non-Fermi-liquid metal.
The inset in Fig.~\ref{d_gamma} indicates $D \propto E^2$ 
for small $E$, which can be understood from the {\it third-order 
nonlinear optical process}, where the correction to 
$\sigma(\nu)|_{E=0}$ is given as 
$\delta\sigma(\nu)\propto\nu{\rm Im}\chi^{(3)}(-\nu; \Omega, -\Omega, \nu)E^2$ 
in terms of the nonlinear optical susceptibility 
$\chi^{(3)}$ (optical Kerr effect) \cite{ButcherCotter}.  

{\it Conclusion}.---We have shown how a 
nonequilibrium steady state is 
determined for photoexcited correlated electrons (in the Falikov-Kimball model) 
by introducing the fermionic bath model.   
This has enabled us to predict the features in the 
optical conductivity that comprise 
kink and dip structures, midgap band, as well as 
negative peaks (gain) as hallmarks of the nonequilibrium.
Future problems include elaborations on 
the coherence of the pump light, dependence on the 
nature of dissipation, 
and robustness of excited states against ultrafast relaxations.

We thank Keiji Saito for a discussion on the bath model.  
The present work is supported in part by a Grant-in-Aid for Scientific Research 
on a Priority Area ``Anomalous quantum materials'' 
from the Japanese Ministry of Education.
N.T. was supported by the Japan Society for the Promotion of Science.

\bibliographystyle{apsrev}
\bibliography{floquet}

\begin{thebibliography}{20}
\bibitem[{dis()}]{dissipation}
\bibinfo{note}{Whether extrinsic dissipation is relevant depends on its
  relaxation time $\Gamma^{-1}$ and the dependence of NESS on $\Gamma$.}
\end{thebibliography}

\end{document}